\documentclass[superscriptaddress,aps,preprintnumbers,amsmath,showpacs,amssymb,prd,nofootinbib,reprint]{revtex4-1}

\usepackage[dvipdfmx]{graphicx}
\usepackage{amsfonts}
\usepackage[mathscr]{eucal}
\usepackage{ascmac}
\usepackage{dcolumn}
\usepackage{bm}
\usepackage[colorlinks=true,linkcolor=blue,citecolor=blue]{hyperref}
\usepackage{hyperref}
\usepackage{color}

\begin{document}


\newcommand{\vev}[1]{ \left\langle {#1} \right\rangle }
\newcommand{\bra}[1]{ \langle {#1} | }
\newcommand{\ket}[1]{ | {#1} \rangle }
\newcommand{\EV}{ \ {\rm eV} }
\newcommand{\KEV}{ \ {\rm keV} }
\newcommand{\MEV}{\  {\rm MeV} }
\newcommand{\GEV}{\  {\rm GeV} }
\newcommand{\TEV}{\  {\rm TeV} }
\newcommand{\1}{\mbox{1}\hspace{-0.25em}\mbox{l}}
\newcommand{\Red}[1]{{\color{red} {#1}}}

\newcommand{\lmk}{\left(}  
\newcommand{\rmk}{\right)}
\newcommand{\lkk}{\left[}  
\newcommand{\rkk}{\right]}
\newcommand{\lhk}{\left \{ }  
\newcommand{\rhk}{\right \} }
\newcommand{\del}{\partial}  
\newcommand{\la}{\left\langle} 
\newcommand{\ra}{\right\rangle}
\newcommand{\half}{\frac{1}{2}}

\newcommand{\bea}{\begin{array}}
\newcommand{\eea}{\end{array}}
\newcommand{\beq}{\begin{eqnarray}}
\newcommand{\eeq}{\end{eqnarray}}

\newcommand{\dd}{\mathrm{d}}
\newcommand{\Mpl}{M_{\rm Pl}}
\newcommand{\mg}{m_{3/2}}
\newcommand{\abs}[1]{\left\vert {#1} \right\vert}
\newcommand{\mphi}{m_{\phi}}
\newcommand{\Hz}{\ {\rm Hz}}
\newcommand{\for}{\quad \text{for }}
\newcommand{\Min}{\text{Min}}
\newcommand{\Max}{\text{Max}}
\newcommand{\Kahler}{K\"{a}hler }
\newcommand{\cphi}{\varphi}
\newcommand{\Tr}{\text{Tr}}
\newcommand{\diag}{{\rm diag}}

\newcommand{\SUf}{SU(3)_{\rm f}}
\newcommand{\Upq}{U(1)_{\rm PQ}}
\newcommand{\Zpq}{Z^{\rm PQ}_3}
\newcommand{\Cpq}{C_{\rm PQ}}
\newcommand{\ubar}{u^c}
\newcommand{\dbar}{d^c}
\newcommand{\ebar}{e^c}
\newcommand{\nubar}{\nu^c}
\newcommand{\Ndw}{N_{\rm DW}}
\newcommand{\Fpq}{F_{\rm PQ}}
\newcommand{\fpq}{v_{\rm PQ}}
\newcommand{\Br}{{\rm Br}}
\newcommand{\Lag}{\mathcal{L}}
\newcommand{\Lqcd}{\Lambda_{\rm QCD}}

\newcommand{\ji}{j_{\rm inf}} 
\newcommand{\jb}{j_{B-L}} 
\newcommand{\M}{M} 
\newcommand{\im}{{\rm Im} }
\newcommand{\re}{{\rm Re} }


\preprint{
TU-1009; 
IPMU 15-0188; 
DESY 15-194
}

\title{
Spontaneous Baryogenesis from Asymmetric Inflaton
}

\author{
Fuminobu Takahashi
}
\affiliation{Department of Physics, Tohoku University, 
Sendai, Miyagi 980-8578, Japan} 
\affiliation{Kavli IPMU (WPI), UTIAS, 
The University of Tokyo, 
Kashiwa, Chiba 277-8583, Japan}

\author{
Masaki Yamada
}
\affiliation{Kavli IPMU (WPI), UTIAS, 
The University of Tokyo, 
Kashiwa, Chiba 277-8583, Japan}
\affiliation{Institute for Cosmic Ray Research, 
The University of Tokyo, 
Kashiwa, Chiba 277-8582, Japan}
\affiliation{
Deutsches Elektronen-Synchrotron DESY, 
22607 Hamburg, Germany
}

\date{\today}

\begin{abstract} 
We propose a variant scenario of spontaneous baryogenesis from asymmetric inflaton
based on current-current interactions between the inflaton and matter fields with a non-zero $B-L$ charge.
When the inflaton starts to oscillate around the minimum after inflation,
it may lead to excitation of a CP-odd component, which induces an effective chemical potential for
the $B-L$ number through the current-current interactions. 
We study concrete inflation models and show that the spontaneous baryogenesis scenario
can be naturally implemented in the chaotic inflation in supergravity.
\end{abstract}


\maketitle


\section{Introduction
\label{sec:introduction}}

The history of the Universe after the Big Bang Nucleosynthesis epoch is
well understood, and it is known as the standard Big Bang cosmology. 
However, it suffers from various initial condition problems
such as the horizon problem, flatness problem, and the origin of density fluctuations. 
In the inflationary paradigm~\cite{Guth:1980zm,Sato:1980yn,Starobinsky:1980te,Brout:1977ix,Kazanas:1980tx}, 
the exponential expansion of the Universe solves these problems. 
On the other hand, in the standard Big Bang cosmology one needs to adopt an initial condition 
such that the amount of baryon number is ten orders of magnitude smaller than the entropy density. 
In the inflationary Universe, any pre-existing baryon asymmetry would be exponentially
diluted, and so, the baryon asymmetry needs to be created after inflation.

The inflation must be connected to the subsequent hot Big Bang phase.
This is naturally realized in the slow-roll inflationary scenario~\cite{Linde:1981mu, Albrecht:1982wi}, where
the inflation is driven by a scalar field called inflaton 
which slowly rolls down the nearly flat potential.
While the Universe is dominated by the potential energy of the inflaton, 
it experiences an exponential expansion.
The inflation ends when the inflaton starts to oscillate around the potential minimum, 
and the potential energy is converted to radiation through the inflaton decay.

Suppose that the inflaton is a part of a complex scalar field with an approximate
U(1) symmetry around the potential minimum.  This is often the case in supersymmetric (SUSY) theories, 
where each chiral multiplet contains a complex scalar. When the inflaton starts to oscillate about
the potential minimum after inflation, it may acquire a non-zero U(1) asymmetry associated with the
inflaton number. (Here the inflaton number refers to the CP-odd component of the excited inflaton quanta.)
Once produced, the inflaton number decreases as $a^{-3}$ due to the expansion of the Universe,
where $a$ is the scale factor. If there are current-current interactions between the inflaton number 
and  the $B-L$ symmetry, the excited CP-odd inflaton quanta induces
an effective chemical potential of the $B-L$ number. 
This leads to the spontaneous baryogenesis if the $B-L$ number is broken in 
thermal plasma~\cite{Cohen:1987vi, Chiba:2003vp, Kusenko:2014lra, D'Agnolo:2015pha},
because the inflaton asymmetry biases the $B-L$ number. As for the $B-L$ breaking, one may introduce
dimension five interactions for the neutrino mass, motivated by the seesaw mechanism~\cite{seesaw}.
Finally the inflaton decays into radiation and reheats the Universe, which connects the inflation to
the hot Big Bang phase.

In this paper,  we consider the spontaneous baryogenesis scenario based on 
the current-current interactions between the inflaton and $B-L$ numbers.
We evaluate the abundance of baryon asymmetry and clarify 
what conditions are needed to explain the observed amount of the baryon asymmetry. 
We also study explicit inflation models and show that the scenario can be naturally
implemented in the chaotic inflation in supergravity. Interestingly, no isocurvature perturbations
are generated in this case, in contrast to the usual spontaneous baryogenesis~\cite{Turner:1988sq}.

The rest of this paper is organized as follows. In the next section we will explain our main idea
about the spontaneous baryogenesis from asymmetric inflaton. In Sec.~\ref{sec:model} we study
concrete inflation models in supergravity. The last section is devoted for discussion and conclusions.

\section{spontaneous baryogenesis from Asymmetric Inflaton
\label{sec:four-Fermi}}
Suppose that inflaton $\phi$ is a complex scalar field 
with an approximate conserved U(1)$_{\rm inf}$ current of 
\beq
 j^{\rm inf}_\mu = 
 2 \im \lmk \phi \del_\mu \phi^* \rmk + \dots. 
\eeq
where $\dots$ represents the other fields carrying nonzero inflaton charges such as the inflatino.
This is the case in many SUSY inflation models because each chiral multiplet contains a complex scalar field. 
In a class of inflation models, the flatness of the inflaton potential is due to the U(1)$_R$ symmetry,
in which case  the inflaton current is identified with the U(1)$_R$ current. 
We assume that the U(1)$_{\rm inf}$ symmetry is only an approximate symmetry at the potential minimum,
but it is generically broken explicitly and in particular, it is generically spontaneously broken during inflation.
When the inflaton starts to oscillate about the potential minimum after inflation, a nonzero inflaton number
may be produced if the U(1)$_{\rm inf}$ is explicitly broken. The inflaton number decreases as $a^{-3}$ if
the breaking term is irrelevant in the vicinity of the potential minimum. 
This is naturally realized in some models as shown in the next section. 
If the inflaton current is coupled to the $B-L$ current, the CP asymmetric inflaton number biases the $B-L$
number.  Thus, spontaneous baryogenesis takes place if $B-L$ number is explicitly broken in plasma.
Some standard model (SM) particles may carry a nonzero inflaton charge to realize the reheating. 
We consider the case that the inflaton charge operator commutes with the $B-L$ charge 
so that we focus on baryon asymmetry generated via the spontaneous baryogenesis. 
Otherwise the baryon asymmetry is directly generated by inflaton dynamics, 
which is the case out of our interest (see, e.g., Ref.~\cite{Kasuya:2014yia} for that case).

To realize spontaneous baryogenesis,  we introduce the following  current-current interaction: 
\beq
- \mathcal{L} = 
 \tilde{G}_F j^{\rm inf}_{\mu}  j^\mu_{B-L}, 
 \label{four-fermi}
\eeq
where $\tilde{G}_F$ is an effective coupling constant with mass dimension minus two.%
\footnote{
In Ref.~\cite{D'Agnolo:2015pha}, 
 a four-Fermi interaction between DM and $B-L$ currents 
with $\tilde{G}_F \sim 1/(10 \TEV)^2$ and the associated spontaneous baryogenesis 
were studied in an asymmetric dark matter model. 
}
Here, $\jb$ is the $B-L$ current,
\beq
 \jb^\mu &=& \sum_i q_i j_i^\mu \\
 j_i^\mu &=& 
 \left\{
 \bea{ll}
 \bar{\psi}_i \gamma^\mu \psi_i 
 &~~ \text{for fermions} 
 \\
 2 \im \lmk \cphi_i \del^\mu \cphi_i^* \rmk
 &~~ \text{for bosons}, 
\eea
 \right.
\eeq
where $q_i$ is the $B-L$ charge of the field $i$. 
Such current-current interactions are generically present in supergravity theories as shown in Sec.~\ref{sec:model}.

After inflation ends, 
the coherent oscillations of the inflaton dominate the energy density of the Universe.
The Friedmann equation implies the following relation: 
\beq
 n_{\rm inf} \simeq \frac{3 \epsilon H^2(t) \Mpl^2}{m_{\rm inf}} 
 \\
 \epsilon \equiv \frac{m_{\rm inf} n_{\rm inf}}{\rho_{\rm inf}}, 
\eeq
where $\Mpl$ ($\simeq 2.4 \times 10^{18} \GEV$) is the reduced Planck mass, 
$m_{\rm inf}$ is the inflaton mass at the potential minimum,
$n_{\rm inf} = (j^{\rm inf})^0$ is the inflaton number density
and 
$\rho_{\rm inf}$ is the energy density of the inflaton. 
Here we have assumed that the only inflaton has a sizable U(1)$_{\rm inf}$ asymmetry.
We define $\epsilon$ ($\le 1$) so that 
it represents the ellipticity of inflaton trajectory in its complex plane. 
As the inflaton number and energy densities are (nearly) spatially homogeneous,
the relevant part of the current-current interaction is given by
\beq
- \mathcal{L} &=&\mu_{B-L} n_{B-L} 
 \\
  \mu_{B-L} &\simeq& 3 \epsilon \tilde{G}_F \Mpl^2 \frac{H^2(t)}{m_{\rm inf}} , 
\eeq
where $n_{B-L} = \jb^0$ is the $B-L$ number density.
This means that the $B-L$ number has 
an effective chemical potential $\mu_{B-L}$. 
The chemical potential biases 
the $B-L$ asymmetry in the chemical equilibrium  as 
\beq
 n_{B-L}^{\rm (eq)} \simeq k \mu_{B-L} T^2 
 \\
 k \equiv \sum_i q_i \frac{g_i}{6}, 
\eeq
where the summation is taken for all particles in the thermal bath 
and 
$g_i$ is the number of spin states but with an extra factor of 2 for bosons. 
In the SM, the coefficient $k$ is $1/2$, 
while in the MSSM it is $1$. 

In order to generate a non-zero $B-L$ asymmetry,
we introduce the following $B-L$ violating interaction
\beq
 \mathcal{L} 
 = 
 \frac{y^2}{2 M_R} (L \tilde{H})^2 + {\rm h.c.}
 \label{dim5}
\eeq
for the light neutrino masses, which is obtained
after integrating out heavy right-handed neutrinos 
in  the seesaw mechanism~\cite{seesaw}. Here
$\tilde H = i \sigma_2 H^*$ is the SU(2) conjugate of the SM Higgs doublet $H$.
Throughout this paper we assume that the right-handed neutrinos are so heavy that
they are not produced from thermal scattering, and so, our scenario is complementary
to  thermal leptogenesis~\cite{Fukugita:1986hr}.
The effective rate of the lepton number violating processes via the above interaction 
is roughly given by~\cite{Buchmuller:2002rq}
\beq
 \Gamma_L \sim \sigma_R T^3 \sim \frac{\bar{m}^2 T^3}{16 \pi v_{\rm ew}^4}, 
\eeq
where $v_{\rm ew} (\simeq 246 \,{\rm GeV})$ is the Higgs vacuum expectation value (VEV)
and 
$\bar{m}^2$ 
is the sum of the left-handed neutrino mass squared. 
We assume $\bar{m}^2$ 
to be of order the atmospheric neutrino mass squared difference, 
$\Delta m_{\rm atm}^2 \simeq 2.4 \times 10^{-3} \EV^2$. (See e.g. 
Refs.~\cite{Forero:2014bxa,Gonzalez-Garcia:2014bfa} for the latest combined-data fit parameters.)
Thus we obtain $\sigma_R \Mpl^2 \simeq 8 \times 10^4$.

Now let us calculate the baryon asymmetry. 
Before the reheating completes, 
the temperature of the plasma is written as 
\beq
 T \simeq \lmk 
 \frac{36 H(t) \Gamma_I \Mpl^2}{g_* \pi^2} 
 \rmk^{1/4}, 
 \label{T}
\eeq
where 
$g_*$ is the effective relativistic degrees of freedom 
and equal to $106.75$ ($228.75$) in the SM (MSSM). 
The inflaton decay rate $\Gamma_I$ is related to the reheating temperature as 
\beq
 T_{\rm RH} 
 \simeq \lmk \frac{90}{ g_* \pi^2} \rmk^{1/4} \sqrt{\Gamma_I \Mpl }. 
 \label{T_RH}
\eeq
Then the ratio between the reaction rate of the $B-L$ violating interaction 
and the Hubble expansion rate is given by 
\beq
 \frac{\Gamma_L}{H} 
 \simeq  
 0.1 \lmk \frac{H_{\rm RH}}{H(t)} \rmk^{1/4} \lmk \frac{T_{RH}}{\Mpl} \rmk 
 \Mpl^2 \sigma_R,
\eeq
where $H_{\rm RH} (=\Gamma_I)$ is the Hubble parameter at the reheating.
When this ratio is larger than unity, 
the $B-L$ asymmetry would reach the equilibrium value. 
However, 
as can be seen from the above expression, the ratio is almost always below unity, 
so that the $B-L$ number at the time of reheating is estimated by integrating the Boltzmann equation:
\beq
 \left. n_{B-L} \right\vert_{\rm RH} \simeq 
 \int_{t_{\rm inf}}^{t_{\rm RH}} \dd t' 
  \frac{H_{\rm RH}^2}{H^2 (t')} 
 \Gamma_L n_{B-L}^{(\rm eq)} (t'),
\eeq
where $t_{\rm RH}$ and $t_{\rm inf}$ represent the cosmic time at the reheating
and the end of inflation, respectively.  Since the integrand is proportional to $T^5$ (i.e., $\propto t^{-5/4}$), 
the baryon abundance is mostly generated 
just after the end of inflation. 
This is in contrast with the ordinary scenario of the spontaneous baryogenesis, 
where the baryon abundance is mostly generated at the time of reheating. 
This is because in our case the effective chemical potential decreases as $\propto H^2 (t)$, 
which is faster than the ordinary case, 
and the bias effect is most efficient just after the end of inflation. 
Thus we can estimate the $B-L$ number at the reheating as
\beq
 \left. n_{B-L} \right\vert_{\rm RH} \simeq 
 \left. 
 \frac{H_{\rm RH}^2}{H^2 (t)} 
 \frac{\Gamma_L}{H(t)} 
 n_{B-L}^{(\rm eq)} (t) \right\vert_{t= t_{\rm inf}}, 
\eeq
Combining those results, 
we obtain the final baryon asymmetry as 
\beq
 Y_b &\equiv& \frac{n_b}{s} 
 \simeq 
  \left. \frac{8 n_f + 4 n_H}{22 n_f  + 13 n_H} T_{\rm RH} \frac{n_{B-L}}{4 H^2 \Mpl^2} \right\vert_{\rm RH}
  \\ 
 &\simeq& 
  0.01\,
 \epsilon 
 (\Mpl^2 \sigma_R)
 (\tilde{G}_F \Mpl^2)
\frac{H(t)^{1/4} T_{\rm RH}^{7/2}}{m_{\rm inf} \Mpl^{11/4}}, 
\label{Y_b}
\eeq
where the pre-factor in the first line is due to the sphaleron effect~\cite{Kuzmin:1985mm}. 
In the SM (MSSM), $n_f = 3$ and $n_H = 1$ ($2$). 
In the second line, we have substituted some parameters, including $g_*$, $n_h$, and $k$, 
and obtain the factor of about $0.01$ for the SM and MSSM. 
The observed baryon abundance of $Y_b^{(\rm obs)} \simeq 8.6 \times 10^{-11}$~\cite{pdg} requires that 
the reheating temperature is as large as
\beq
 T_{RH} 
&\simeq& 3 \times 10^{13} \GEV 
\epsilon^{-2/7} 
 \lmk \tilde{G}_F \Mpl^2 \rmk^{-2/7} 
  \nonumber
  \\
  && \times \lmk \frac{m_{\rm inf}}{10^{13} \GEV} \rmk^{2/7} 
  \lmk \frac{H(t)}{10^{13} \GEV} \rmk^{-1/14}. 
  \label{result}
\eeq

Lastly let us comment on the washout effect. 
Around the time of reheating, 
the generated lepton asymmetry is partially washed out 
due to the inverse processes. 
Calculating the Boltzmann equation, 
we obtain a washout factor such as~\cite{Kusenko:2014lra}
\beq
 \Delta_w \simeq 
 \exp \lkk 
 -0.7 \lmk \frac{T_{\rm RH}}{10^{13} \GEV} \rmk
 \rkk. 
 \label{washout}
\eeq
This implies that 
the reheating temperature cannot be much larger than $10^{13} \GEV$ 
to avoid the washout effect due to the inverse processes. 
Taking into account the washout factor, 
we find that the resulting baryon asymmetry has a peak around the reheating temperature of $5 \times 10^{13}$, 
where the result of Eq.~(\ref{Y_b}) is overestimated by a factor of $5$. 
Thus we may assume $(\tilde{G}_F \Mpl^2) \simeq 5$ to explain the observed baryon asymmetry.

\section{Inflation Models 
\label{sec:model}}

In this section, we consider concrete models of inflation in supergravity 
to see if the scenario in the previous section can be realized in realistic inflation models. 
Hereafter, we adopt the Planck units ($\Mpl = 1$).

In supergravity,  the Lagrangian is constructed to be invariant in terms of supergravity transformation. 
The relevant part of the Lagrangian comes from kinetic terms such as 
\beq
 \mathcal{L} &=& 
  K_{i j^*} \del_\mu \cphi^i \del^\mu \cphi^{* j}
 + i K_{i j^*} \bar{\chi}^j \bar{\sigma}^\mu \tilde{\mathcal{D}}_\mu \chi^i, 
 \\
 \tilde{\mathcal{D}}_\mu \chi^i &\equiv& \del_\mu \chi^i + K^{i m^*} K_{k m^* l} \del_\mu \cphi^l \chi^k \nonumber
 \\
 &&- \frac{1}{4} \lmk K_j \del_\mu \cphi^j - K_{j^*} \del_\mu \cphi^{*j} \rmk \chi^i + \dots, 
\eeq
where scalar and fermionic components of chiral superfields are denoted by  $\cphi_i$ and $\chi_i$, respectively, and
$K$ is the \Kahler potential, 
$K^{i j^*} \equiv (K_{i j^*})^{-1}$, 
and the subscripts represent derivatives with respect to the corresponding field, 
such as $K_{i j^*} \equiv \del^2 K / \del \cphi^i \del \cphi^{*j}$. 
Note that the total Lagrangian is real.

\subsection{F-term hybrid inflation
\label{sec:hybrid inflation}}

Suppose that there is a \Kahler potential of 
\beq
 K = \abs{\phi}^2 + c \abs{\phi}^2 \abs{\chi}^2 + \abs{\chi}^2, 
\eeq
where the scalar component of $\phi$ is inflaton 
and $\chi$ is a $B-L$ charged field. 
Neglecting the fermionic component of $\phi$ 
and denoting the scalar and fermionic components of $\chi$ as $\tilde{\chi}$ and $\chi$, respectively, 
we obtain the desirable interaction term such as 
\beq
 \mathcal{L}_{\rm int}  
 = 
 \im \lmk \phi \del_\mu \phi^* \rmk 
 \lkk 
 2 c  \im \lmk \tilde{\chi} \del^\mu \tilde{\chi}^* \rmk 
 + \lmk c - \frac{1}{2}  \rmk
 \bar{\chi} \bar{\sigma}^\mu \chi 
 \rkk
\nonumber
\\
\eeq
where we have assumed $c \abs{\phi}^2 \ll 1$ 
and have rescaled the fields $\tilde{\chi}$ and $\chi$ to make their kinetic term canonical. 
These interactions are nothing but the current-current interaction of the form (\ref{four-fermi}) 
with $(\tilde{G}_F \Mpl^2) \approx c$.

The above calculation can be applied to, e.g., 
the F-term hybrid inflation model with the superpotential~\cite{Copeland:1994vg, Dvali:1994ms}
\beq
 W = \lambda \phi \lmk \psi \bar{\psi} - \frac{v^2}{2} \rmk, 
\eeq
where $\psi$ and $\bar{\psi}$ are waterfall fields, and $\phi$ is the inflaton with U(1)$_R$ charge $2$.
Note that in this case the inflaton current is nothing but the R-current. The U(1)$_R$ symmetry is
necessarily broken by a constant term of the superpotential, $W_0 \simeq m_{3/2}$, 
which is required for realizing  a vanishingly small cosmological constant. As a result, the inflaton potential
receives a linear term  potential in proportion to the gravitino 
mass, $m_{3/2}$, which modifies the inflaton dynamics~\cite{Buchmuller:2000zm,Nakayama:2010xf}.
In particular, the angular motion of $\phi$, i.e., the inflaton number, 
is induced during  inflation~\cite{Nakayama:2010xf}. The effect of the angular motion on the density perturbations
was recently studied in detail in Ref.~\cite{Buchmuller:2014epa}.
The dynamics of inflaton during inflation is mainly determined by the Coleman-Weinberg potential 
and the linear term of inflaton. 
The following parameter $\xi$ measures the relative importance of the two contributions to 
the slope of the potential~\cite{Buchmuller:2014epa}: 
\beq
  \xi \equiv \frac{2^{9/2} \pi^2}{\kappa^3 \ln 2} \frac{m_{3/2}}{v} ~~ \lmk \le 1 \rmk. 
\eeq
Thus we obtain the ellipticity parameter of inflaton trajectory after inflation as 
\beq
 \epsilon \simeq \frac{n_{\rm inf}}{ m_{\rm inf} v^2} \simeq \frac{H_{\rm inf}}{m_{\rm inf}} \xi, 
\eeq
where the factor of $H_{\rm inf} / m_{\rm inf}$ comes from 
the difference of the time scale of inflaton dynamics between the eras during and after inflation.

When we consider the gravitino mass of order $100 \TEV$, 
the spectral index can be consistent with the observed value 
for the case of $\lambda \simeq 3 \times 10^{-3}$ and $v \simeq 4 \times 10^{15} \GEV$
for the final phase value of order unity~\cite{Buchmuller:2014epa}. 
This implies that 
the mass of inflaton and the Hubble parameter just after inflation 
are given by 
\beq
 m_{\rm inf} &=& \lambda v \simeq 1.2 \times 10^{13} \GEV 
 \\
 H_{\rm inf} &\simeq& \frac{\lambda v^2}{2 \sqrt{3} \Mpl} \simeq 5.7 \times 10^9 \GEV. 
\eeq
From Eq.~(\ref{result}), 
we find that the reheating temperature needs to be as high as 
\beq
 6 \times 10^{13} \GEV 
\epsilon^{-2/7} 
 \lmk \tilde{G}_F \Mpl^2 \rmk^{-2/7}, 
\eeq
to explain the observed baryon asymmetry. 
Note that for the above parameters 
the ellipticity parameter $\epsilon$ is of order $10^{-4}$, 
so that we need $\tilde{G}_F \Mpl^2 \simeq 10^{4}$ (i.e., $c \simeq 10^{4}$) 
to explain the observed baryon asymmetry. 
In addition, 
the reheating temperature cannot be as high as $10^{13} \GEV$ in the hybrid inflation model. 
This is because 
couplings between inflaton and other particles have to be suppressed by $\kappa$ ($\ll 1$) 
in order not to affect the inflaton potential 
and 
the nonperturbative enhancement of decay process called preheating is suppressed after inflation 
in the case of the rotating inflaton~\cite{Chacko:2002wr}. 
Thus,  we conclude that 
both 
the ellipticity parameter $\epsilon$ and 
the reheating temperature are too small to account for the observed baryon abundance 
in the hybrid inflation model.

\subsection{Chaotic inflation
\label{sec:chaotic inflation}}

As another example, let us focus on a chaotic inflation model proposed in Ref.~\cite{Kawasaki:2000yn},
where the \Kahler potential respects
a shift symmetry of the inflaton field,
\beq
\phi \to \phi + i \alpha,
\eeq
where $\alpha$ is the transformation parameter.
The \Kahler potential is given by
\beq
 K &=& c' (\phi + \phi^*) + \frac{1}{2} \lmk \phi + \phi^* \rmk^2 + \abs{X}^2 \nonumber\\
 &+& \abs{\chi}^2 + \frac{c}{2} \lmk \phi + \phi^* \rmk^2 \abs{\chi}^2, 
\eeq
where $X$ is a stabilizer field.  The relevant interactions of the Lagrangian are then given by 
\beq
 &&\mathcal{L}_{\rm int}  
 = 
 \frac{c'}{2} \del_\mu \im [ \phi ] 
 \bar{\chi} \bar{\sigma}^\mu \chi 
 \nonumber\\
 &&- 2 \re [ \phi ] \del_\mu \im [ \phi ] 
  \lkk 
 2 c  \im \lmk \tilde{\chi} \del^\mu \tilde{\chi}^* \rmk 
 + \lmk c - \frac{1}{2}  \rmk
 \bar{\chi} \bar{\sigma}^\mu \chi 
 \rkk, 
 \nonumber\\
\eeq
where we have rescaled the fields to obtain the canonical kinetic terms.
The shift symmetry is assumed to be explicitly broken in the superpotential,
\beq
 W = m_{\rm inf} \phi X,
\eeq
where the R-charge assignment is $R[X]=2$ and $R[\phi]=0$.
When the constant $c'$ is nonzero, 
the real part of $\phi$ has a VEV of order $c'$ during inflation. 
Thus 
the inflaton 
starts to rotate in the complex plane after inflation such as 
\beq
 \re [\phi] \approx c' |\phi| \sin m_{\rm inf} t \\
 \im [\phi] \approx |\phi| \cos m_{\rm inf} t. 
\eeq
Note here that, in addition to the U(1)$_R$ symmetry,
 the scalar potential has another (approximate) global U(1) symmetry for which
 $\phi$ and $X$ have  the same magnitude charge but opposite sign.
This implies that the inflaton number is induced after inflation: 
\beq
 \re [\phi] \del_0 \im [\phi] \simeq c' m_{\rm inf} \abs{\phi}^2 \simeq c' \frac{H^2}{m_{\rm inf}}. 
\eeq
That is, the ellipticity parameter is given by $\epsilon \simeq c'$, 
which is expected to be of order unity, 
and the effective coupling is given by $(\tilde{G}_F \Mpl^2) \approx c$.

The mass of inflaton is determined by the COBE normalisation such as $m_{\rm inf} \simeq 10^{13} \GEV$, 
which implies that the Hubble parameter just after inflation is given by $6 \times 10^{12} \GEV$. 
Thus the observed baryon asymmetry can be explained when 
the reheating temperature is as large as 
\beq
 T_{\rm RH} \simeq 
 3 \times 10^{13 } \GEV 
 \lmk \tilde{G}_F \Mpl^2 \rmk^{-2/7}, 
\eeq
where we assume $\epsilon = 1$. 
Note that $(\tilde{G}_F \Mpl^2 )$ ($\approx c$) is expected to be of order unity. 
To obtain such high reheating temperature, 
we introduce a superpotential of~\cite{Kawasaki:2000yn}
\beq
 W_{\rm RH} = y X H_u H_d, 
\eeq
where $H_u$ and $H_d$ are a pair of Higgs doublets in the MSSM sector. 
This leads to the coupling between the inflaton and the Higgs doublets as $\mathcal{L}_{\rm RH} \sim y m_{\rm inf} 
\phi H_u H_d$, 
leading to  the reheating temperature given by%
\footnote{
The coupling constant $y$ should be smaller than of order $m_{\rm inf} \phi \approx 10^{-5}$ 
so that the $H_u H_d$ VEV does not cancel the F-term of $X$~\cite{Nakayama:2013txa}. 
This problem can be avoided, e.g., when 
we introduce the higher dimensional superpotential of $(H_u H_d)^2$ 
to prevent the $H_u H_d$ direction from obtaining a large VEV during inflation. 
}
\beq
 T_{\rm RH} \sim 10^{13} \GEV \lmk \frac{y}{0.1} \rmk \lmk \frac{m_{\rm inf}}{10^{13} \GEV} \rmk^{1/2}. 
\eeq
Therefore, 
in the chaotic inflation model 
our scenario of spontaneous baryogenesis works naturally 
and can explain the observed baryon abundance without any fine-tunings of the parameters.

\section{discussion and conclusions
\label{sec:conclusion}}

In this paper we have proposed a scenario of spontaneous baryogenesis 
from asymmetric inflaton via current-current interactions 
between the inflaton and $B-L$ numbers. 
Such interactions are naturally present in supergravity theories. 
The CP asymmetric part of the inflaton number  induces an effective 
chemical potential of $B-L$ number, which biases the $B-L$ asymmetry in the equilibrium state. 
If the $B-L$ number is  broken in the plasma, a non-zero $B-L$ asymmetry is induced.
We have shown that the observed baryon abundance 
can be explained by this mechanism 
when the reheating temperature is as large as $10^{13} \GEV$. 
We have also studied concrete inflation models in supergravity to see if
the above mechanism can be implemented, and we have found that
our mechanism naturally explain 
the observed baryon abundance without fine-tunings in the chaotic inflation model.

The quadratic chaotic inflation predicts a rather large tensor-to-scalar ratio $r$,
which is now strongly disfavored by the CMB observations~\cite{Ade:2015lrj,Ade:2015tva}. 
The tensor-to-scalar ratio can be easily reduced by introducing higher order terms of the inflaton,
in which case the inflaton potential is given by a polynomial 
function~\cite{Nakayama:2013jka,Nakayama:2013txa}. Our spontaneous baryogenesis 
scenario works similarly in this case.

It is known that  the spontaneous baryogenesis in the slow-roll regime 
generically leads to  sizable isocurvature perturbations~\cite{Turner:1988sq}.
This makes the spontaneous baryogenesis incompatible with high-scale inflation.
In the case of chaotic inflation studied in the previous section, both the real component of $\phi$
as well as the stabilizer field can have a mass of order the Hubble parameter during inflation, and so
there is no light degrees of freedom during inflation other than the inflaton.
Thus, no isocurvature perturbations are induced.

In general, high reheating temperature $T_{RH} \sim 10^{13}$\,GeV is required for
successful baryogenesis in our scenario, and gravitinos are copiously produced 
from thermal scattering. On the other hand, non-thermal gravitino production can be
suppressed at such high reheating temperature~\cite{Kawasaki:2006gs,Endo:2007ih,Endo:2007sz}. 
If the gravitino is lighter than of order $100\TEV$, 
its decay destroys light elements, altering their abundances in contradiction with observations.
If it is heavier than $100 \TEV$, 
it decays before the BBN epoch 
but the LSPs may be overproduced. 
In order to avoid the overproduction of the LSPs, 
we may assume that the R-parity is violated  and the LSP decays before the BBN epoch. 
In this case, we require another dark matter candidate such as axion.

\vspace{1cm}

%
\section*{Acknowledgments}
M.Y. thanks W. Buchm\"{u}ller for kind hospitality at DESY where this work was finished. 
This work is supported by MEXT KAKENHI Grant Numbers 15H05889 and 23104008 (F.T.),
JSPS KAKENHI Grant Numbers 24740135,  26247042, and 26287039 (F.T.), 
JSPS Research Fellowships for Young Scientists (No.25.8715 (M.Y.)), 
World Premier International Research Center Initiative (WPI Initiative), MEXT, Japan (F.T. and M.Y.),
and the Program for the Leading Graduate Schools, MEXT, Japan (M.Y.).
%

\vspace{1cm}




\begin{thebibliography}{90}

  
\bibitem{Guth:1980zm} 
  A.~H.~Guth,
  Phys.\ Rev.\ D {\bf 23}, 347 (1981).
  
\bibitem{Brout:1977ix} 
  R.~Brout, F.~Englert and E.~Gunzig,
  Annals Phys.\  {\bf 115}, 78 (1978).
  
\bibitem{Kazanas:1980tx} 
  D.~Kazanas,
  Astrophys.\ J.\  {\bf 241}, L59 (1980).
  
\bibitem{Starobinsky:1980te} 
  A.~A.~Starobinsky,
  Phys.\ Lett.\ B {\bf 91}, 99 (1980).
  
\bibitem{Sato:1980yn} 
  K.~Sato,
  Mon.\ Not.\ Roy.\ Astron.\ Soc.\  {\bf 195}, 467 (1981).
  
  
\bibitem{Linde:1981mu} 
  A.~D.~Linde,
  Phys.\ Lett.\ B {\bf 108}, 389 (1982).
  
\bibitem{Albrecht:1982wi} 
  A.~Albrecht and P.~J.~Steinhardt,
  Phys.\ Rev.\ Lett.\  {\bf 48}, 1220 (1982).
  
\bibitem{Cohen:1987vi} 
  A.~G.~Cohen and D.~B.~Kaplan,
  Phys.\ Lett.\ B {\bf 199}, 251 (1987);
  Nucl.\ Phys.\ B {\bf 308}, 913 (1988);
  %
  M.~Dine, P.~Huet, R.~L.~Singleton, Jr and L.~Susskind,
  Phys.\ Lett.\ B {\bf 257}, 351 (1991);
  %
  A.~G.~Cohen, D.~B.~Kaplan and A.~E.~Nelson,
  Phys.\ Lett.\ B {\bf 263}, 86 (1991). 

  
  

\bibitem{Chiba:2003vp} 
  T.~Chiba, F.~Takahashi and M.~Yamaguchi,
  Phys.\ Rev.\ Lett.\  {\bf 92}, 011301 (2004)
  [Phys.\ Rev.\ Lett.\  {\bf 114}, no. 20, 209901 (2015)]
  [hep-ph/0304102]; 
%
  F.~Takahashi and M.~Yamaguchi,
  Phys.\ Rev.\ D {\bf 69}, 083506 (2004)
  [hep-ph/0308173].
  
  
\bibitem{Kusenko:2014lra} 
  A.~Kusenko, L.~Pearce and L.~Yang,
  Phys.\ Rev.\ Lett.\  {\bf 114}, no. 6, 061302 (2015)
  [arXiv:1410.0722 [hep-ph]]; 
%
  A.~Kusenko, K.~Schmitz and T.~T.~Yanagida,
  Phys.\ Rev.\ Lett.\  {\bf 115}, no. 1, 011302 (2015)
  [arXiv:1412.2043 [hep-ph]];
  %
  M.~Ibe and K.~Kaneta,
  Phys.\ Rev.\ D {\bf 92}, no. 3, 035019 (2015)
  [arXiv:1504.04125 [hep-ph]].
  R.~Daido, N.~Kitajima and F.~Takahashi,
  arXiv:1504.07917 [hep-ph];
%
  F.~Takahashi and M.~Yamada,
  JCAP {\bf 1510}, no. 10, 010 (2015)
  [arXiv:1507.06387 [hep-ph]].

\bibitem{D'Agnolo:2015pha} 
  R.~T.~D'Agnolo and A.~Hook,
  Phys.\ Rev.\ D {\bf 91}, no. 11, 115020 (2015)
  [arXiv:1504.00361 [hep-ph]].



  
\bibitem{seesaw} 
T.~Yanagida, Conf.\ Proc.\ C {\bf 7902131}, 95 (1979); 
T.~Yanagida, 
Prog.\ Theor.\ Phys.\ \textbf{64}, 1103 (1980); 
M.~Gell-Mann, P.~Ramond and R.~Slansky, Conf.\ Proc.\ C {\bf 790927}, 315 (1979); 
see also P.~Minkowski,
Phys.\ Lett.\ B \textbf{67}, 421 (1977).



\bibitem{Turner:1988sq} 
  M.~S.~Turner, A.~G.~Cohen and D.~B.~Kaplan,
  Phys.\ Lett.\ B {\bf 216}, 20 (1989).


\bibitem{Kasuya:2014yia} 
  S.~Kasuya and F.~Takahashi,
  Phys.\ Lett.\ B {\bf 736}, 526 (2014)
  [arXiv:1405.4125 [hep-ph]].

   
\bibitem{Fukugita:1986hr} 
  M.~Fukugita and T.~Yanagida,
  Phys.\ Lett.\ B {\bf 174}, 45 (1986).
  
  

\bibitem{Buchmuller:2002rq} 
  W.~Buchmuller, P.~Di Bari and M.~Plumacher,
  Nucl.\ Phys.\ B {\bf 643}, 367 (2002)
  [Nucl.\ Phys.\ B {\bf 793}, 362 (2008)]
  [hep-ph/0205349].
  
\bibitem{Forero:2014bxa} 
  D.~V.~Forero, M.~Tortola and J.~W.~F.~Valle,
  Phys.\ Rev.\ D {\bf 90}, 093006 (2014)
  [arXiv:1405.7540 [hep-ph]].

\bibitem{Gonzalez-Garcia:2014bfa} 
  M.~C.~Gonzalez-Garcia, M.~Maltoni and T.~Schwetz,
  JHEP {\bf 1411}, 052 (2014)
  [arXiv:1409.5439 [hep-ph]].



\bibitem{Kuzmin:1985mm} 
  V.~A.~Kuzmin, V.~A.~Rubakov and M.~E.~Shaposhnikov,
  Phys.\ Lett.\ B {\bf 155}, 36 (1985).



\bibitem{pdg} 
  K.~A.~Olive {\it et al.} [Particle Data Group Collaboration],
  Chin.\ Phys.\ C {\bf 38}, 090001 (2014).




\bibitem{Copeland:1994vg} 
  E.~J.~Copeland, A.~R.~Liddle, D.~H.~Lyth, E.~D.~Stewart and D.~Wands,
  Phys.\ Rev.\ D {\bf 49}, 6410 (1994)
  [astro-ph/9401011].



\bibitem{Dvali:1994ms} 
  G.~R.~Dvali, Q.~Shafi and R.~K.~Schaefer,
  Phys.\ Rev.\ Lett.\  {\bf 73}, 1886 (1994)
  [hep-ph/9406319].

\bibitem{Buchmuller:2000zm} 
  W.~Buchmuller, L.~Covi and D.~Delepine,
  Phys.\ Lett.\ B {\bf 491}, 183 (2000)
  [hep-ph/0006168].

\bibitem{Nakayama:2010xf} 
  K.~Nakayama, F.~Takahashi and T.~T.~Yanagida,
  JCAP {\bf 1012}, 010 (2010)
  [arXiv:1007.5152 [hep-ph]].

\bibitem{Buchmuller:2014epa} 
  W.~Buchmuller, V.~Domcke, K.~Kamada and K.~Schmitz,
  JCAP {\bf 1407}, 054 (2014)
  [arXiv:1404.1832 [hep-ph]].
  
  
\bibitem{Chacko:2002wr} 
  Z.~Chacko, H.~Murayama and M.~Perelstein,
  Phys.\ Rev.\ D {\bf 68}, 063515 (2003)
  [hep-ph/0211369].
  
\bibitem{Kawasaki:2000yn} 
  M.~Kawasaki, M.~Yamaguchi and T.~Yanagida,
  Phys.\ Rev.\ Lett.\  {\bf 85}, 3572 (2000)
  [hep-ph/0004243].



\bibitem{Ade:2015lrj} 
  P.~A.~R.~Ade {\it et al.}  [Planck Collaboration],
  arXiv:1502.02114 [astro-ph.CO].
  
\bibitem{Ade:2015tva} 
  P.~A.~R.~Ade {\it et al.} [BICEP2 and Planck Collaborations],
  Phys.\ Rev.\ Lett.\  {\bf 114}, 101301 (2015)
  [arXiv:1502.00612 [astro-ph.CO]].

\bibitem{Nakayama:2013jka} 
  K.~Nakayama, F.~Takahashi and T.~T.~Yanagida,
  Phys.\ Lett.\ B {\bf 725}, 111 (2013)
  [arXiv:1303.7315 [hep-ph]].

\bibitem{Nakayama:2013txa} 
  K.~Nakayama, F.~Takahashi and T.~T.~Yanagida,
  JCAP {\bf 1308}, 038 (2013)
  [arXiv:1305.5099 [hep-ph]].
  
  
\bibitem{Kawasaki:2006gs} 
  M.~Kawasaki, F.~Takahashi and T.~T.~Yanagida,
  Phys.\ Lett.\ B {\bf 638}, 8 (2006)
  [hep-ph/0603265].

\bibitem{Endo:2007ih} 
  M.~Endo, F.~Takahashi and T.~T.~Yanagida,
  Phys.\ Lett.\ B {\bf 658}, 236 (2008)
  [hep-ph/0701042].

\bibitem{Endo:2007sz} 
  M.~Endo, F.~Takahashi and T.~T.~Yanagida,
  Phys.\ Rev.\ D {\bf 76}, 083509 (2007)
  [arXiv:0706.0986 [hep-ph]].
  
\end{thebibliography}
\end{document}